\documentclass{iaus}
\usepackage{graphicx}

\title[Spitzer View of Lyman Break Galaxies] 
{Spitzer View of Lyman Break Galaxies}
\author[G.E. Magdis \& D. Rigopoulou]   
{Georgios E. Magdis$^1$%
  \thanks{Present address: University of Oxford,DWB, Keble Rd, Oxford, OX1 3RH, U.K.},
 Dimitra Rigopoulou$^1$}

\affiliation{$^1$Department of Astrophysics, University of Oxford,
Keble Rd, Oxford, OX1 3RH, U.K. \break email: gem@astro.ox.ac.uk, dar@astro.ox.ac.uk\\[\affilskip]}

\pubyear{2004}
\volume{xxx}  
\pagerange{119--126}
\date{?? and in revised form ??}
\jname{Proceedings Title IAU Symposium}
\editors{A.C. Editor, B.D. Editor \& C.E. Editor, eds.}
\begin{document}

\maketitle

\begin{abstract}
Using a combination of deep MID-IR observations obtained by IRAC, MIPS and IRS on board Spitzer we investigate the MID-IR properties of Lyman Break Galaxies (LBGs) at z$\sim$3,
 establish a better understanding of their nature and attempt a complete characterisation of the population. With deep mid-infrared and optical 
observations of $\sim$1000 LBGs covered by IRAC/MIPS and from the ground respectively, we extend  the 
spectral energy distributions (SEDs) of the LBGs to mid-infrared. Spitzer data reveal for the first time that the mid-infrared properties of the population are inhomogeneous ranging from those with marginal IRAC detections to those with bright rest-frame near-infrared colors and those detected at 24$\mu$m MIPS band revealing the newly discovered population of the Infrared Luminous Lyman Break Galaxies (ILLBGs). To investigate this diversity, we examine the photometric properties of the population and we use stellar population synthesis models to probe the stellar
content of these galaxies. We find that a fraction of LBGs have very red colors and large estimated stellar
masses $M_{\ast}$$>$5$\times$$10^{10}$$M_{\odot}$. We discuss the link between these LBGs
and submm-luminous galaxies and we report the detection of rest frame 6.2 and 7.7 $\mu$m emission features arising
from Polycyclic Aromatic Hydrocarbons (PAH) in the Spitzer/IRS spectrum of
an infrared-luminous Lyman break galaxy at z=3.01.

\end{abstract}

\firstsection 
\section{Introduction}

Observation and study of high-redshift galaxies is essential to constrain the history of galaxy evolution and give us a systematic and quantitative 
picture of galaxies in the early universe, an epoch of rigorous star and galaxy formation. Large samples of high-z galaxies 
that have recently become available, play a key role to that direction and have revealed a zoo of different galaxy populations at z. 
There are various techniques for detecting high-z galaxies involving observations in wavelengths that span from optical to far-IR. 
Among the various methods the Lyman break dropout technique (\cite{Steidel93}), sensitive to the presence of the 912{\AA} break, 
is designed to select z$\sim$ 3 galaxies. LBGs constitute at the moment the largest galaxy population at z$\sim$3 (\cite{Steidel03}). 
With observations spanning from X-rays (eg. \cite{Nan02}) to near-infrared (\cite{Shapley03}) there has been a considerable progress into 
understanding the nature of population, but to fully characterize their properties 
(such as stellar mass, dust content, link to other z$\sim$3 populations) observations of longer wavelengths are required.

With the advent of Spitzer Space Telescope (\cite{Werner04}) we have access to longer wavelengths. IRAC bands (3.6, 4.5, 5.8, 8.0$\mu$m) are crucial as they trace 
the rest-frame near infrared luminosities for galaxies at 0.5$<$z$<$5, (where the bulk of 
the stellar mass of a galaxy radiates) while MIPS (24, 70, 160 $\mu$m) and IRS (5.3--40$\mu$m) provide an insight 
into the interstellar medium of the population as they are sensitive to PAH features and dust re-radiation.    

In this study we use IRAC and MIPS data covering $\sim$1000 and 244 LBGs respectively, lying on the fields Q1422+2309 (Q1422), DSF2237a,b (DSF), Q2233+1341 (Q2233), SSA22a,b
(SSA22), B20902+34 (B0902), QSOHS1700+6416 (Q1700), Extended Groth Strip (EGS) and Hubble Deep Field North (HDFN). 
Those LBGs have previously beeen identified from their optical colours by \cite{Steidel03}. 
In section 2 we search for mid-infrared counterparts of the LBGs, extend their SEDs to mid-infrared 
and investigate their mid-infrared colours as well as their physical properties such as stellar mass and dust content. 
In section 3 we examine the possible link between the IRAC/MIPS bright LBGs and the SMGs while in Section 
4 using data obtained by IRS, we report the detection of PAH features arising from the mid-inrared spectrum 
of an ILLBG at z=3.01. In Section 5 we summurize the results of this Spitzer view of LBGs. 

\section{Mid-infrared Properties of LBGs}\label{sec:greenfun}

\subsection{The Spectral Energy Distridution of LBGs}

In a sample of 768 LBGs lying in the fields of our study (excluding EGS) we report the detection of 457,
456, 151 and 158 at 3.6, 4.5, 5.8 and 8.0$\mu$m IRAC bands respectively. Source extraction, photometric analysis, number counts and mid-infrared identification of LBGs  are 
discussed in detail by Magdis et. al 2007 (in prep.).  We extend the spectral energy distribution of the LBGs 
to rest frame near-infrared and improve dramatically our understanding of the nature of LBGs. 
Figure 1 shows the rest-UV/optical/near-infrared SEDs of all LBGs of the current
sample with confirmed spectroscopic redshift. UV/optical data are obtained from Steidel
et al. 2003, while IRAC data come from the present work. While the rest-UV/optical
show little variation (2--3 magnitudes), the rest frame near infrared colour spread over 6
magnitudes. The addition of IRAC bands reveals for the first time that LBGs display a
variety of colors and their rest-near-infrared properties are rather inhomogeneous, ranging
from :
\begin{itemize}
\item Those that are bright in IRAC bands and exhibit $R-[3.6]>1.5$ colours.
Their SEDs are rising steeply towards
    longer wavelengths and
    based on their R$-$[3.6] we call them $''$\textsl{red}$''$ LBGs-z, to
\item Those that are faint or not detected at all in IRAC bands with
    $R-[3.6]<1.5$ colour. Their SEDs are rather flat from the
far-UV to the NIR with marginal IRAC detections and as they exhibit bluer
R$-$[3.6] colours we call them $''$\textsl{blue}$''$ LBGs-z.

\end{itemize}
\begin{figure}[!h]
\centering
 \includegraphics[width=4cm,height=8cm,angle=-90]{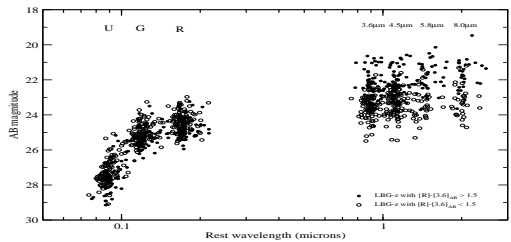}
  \caption {\small{Spectral energy distribution for LBGs with spectroscopic redshift and classified as galaxies. Empty circle represent LBGs with $R-[3.6]<1.5$, while filled circle\
s represent LBGs with $R-[3.6]<1.5$, revealing that the rest-near-infrared of the population displays a wide range of colours}}
\end{figure}

\subsection{The 8$\mu$m sample, Stellar Masses and Dust Content}
From the whole sample of the detected at 3.6$\mu$m and 4.5$\mu$m LBGs, a fraction of about 40\% 
are detected in longer wavelengths, creating the the sample of the 8$\mu$m LBGs. We find that 
LBGs with 8$\mu$m counterpart temd to exhibit redder $<R-[3.6]>$ colors when compared to those that 
are not detected at 8$\mu$m with a mean value of 1.91$\pm$0.16 and 1.02$\pm$0.26 respectively. The significance of the 
8$\mu$m sample is that for z$\sim$3, 8$\mu$m correspond to K rest-frame, sensitive to the bulk of the stellar emission of a galaxy and not only to the young population
of a recent star-forming event. In the first study of Spitzer detection of LBGs in the EGS field (covering 244 LBGs), \cite{Rig06} (R06 hereafter), using stellar 
synthesis population models generated with BC03, found that LBGs with 8$\mu$m are more massive dustier and relatively older when compared to the rest of the population. 
Preliminary results from our study (Magdis et. al in prep.), using both BC03 and CB07 (including an observationaly calibrated AGB phase) are in excellent agreement with 
R06. We find that LBGs with 8$mu$m counterpart have estimated stellar masses of $\sim$ 8$\times$$10^{10}$ and dust content of E(B-V)$\sim$0.25 while the mean values for the 
rest of the population are 2.7$\times$$10^{10}$ and 0.18 respectively.

\section{ILLGS and the Link to SMGs}

From a sample of 244 LBGs in the EGS covered by MIPS, a fraction of about $\sim$ 5\% where detected at 24$\mu$m creating the sample of the Infrared Luminous Lyman 
Break Galaxies (ILLBGs) sample (\cite{H05}). The detection of those LBGs at 24$\mu$m indicate the existance of significant amount of dust.
 R06 found the these LBGs have estimated stellar masses $M_{\ast}>10^{11}M_{\odot}$, ages 
$>$ 1000 Myrs and with exctinction that 
varies around A$_{V}$$\sim$0.2. These masses are similar to those of SMG as discussed by \cite{Borys05}. Also R06 found that ILLBGs show extreme R-K $>$ 3 colors similar to 
the colors of SMGs examined by \cite{Chap00}. Furthermore, ILLBGs and SMGs appear to have
 similar mid-infrared R-[3.6] and [8.0]-[24] colors (R06, \cite{Ashby06}). As these two population share many properties in common, it can be 
suggested that there must be a link between them. A possible scenario is one where SMGs and LBG form a continuum of objects with SMGs being the most dusty and most 
star-forming LBGs. In that case, ILLBGs should be at least detected in the submm bands. Therefore, we carried out submm observation of several 
ILLBGs with IRAM-MAMBO and we report the detection of 2 ILLBGs in the submm bands (Westphal-C47 (4$\sigma$) and Westphal-M30 (3$\sigma$)). 
The data for the rest of the ILLBGs have not been analysed yet.     

\section{IRS Spectroscopy of ILLBGs}
To further investigate the dust properties of the LBGs, we carried out follow up IRS mid-infrared spectroscopy of a number of ILLBGs from our sample. As preliminary result
 of this project, we report the detection of strong PAH features arising from the spectrum of the EGS21 ILLBG galaxy at z=3.01 (Figure 2) (\cite{H07}). This is currently the highest
redshift galaxy where these PAH emission features have been detected. The total infrared luminosity inferred from the MIPS 24$\mu$m and radio flux density
is 2$\times$$10^{13} L_{\odot}$, which qualifies this object as a so-called hyper-luminous infrared galaxy (HyLIRG). 
Given the strong  PAH emission features and the lack of any evidence for an AGN, we conclude that
star-formation dominates  the emission from this z=3.01 ILLBG. 

\section{Conclusions}

The advent of Spitzer has dramatically improved our understanding of the LBGs. Using data obtained by IRAC/MIPS/IRS on board Spitzer we have reached the 
following conclusions for the population of the LBGS:

\begin{itemize}
\item IRAC colors have revealed the diversity of LBGs ranging from those with marginal detection in IRAC bands and R-[3.6]$<$1.5 colors, to those that have 
bright in IRAC bands and exhibit R-[3.6]$>$1.5 colors.
\item LBGs detected at 8$\mu$m have redder R-[3.6] colors and on average are more massive, suffer more obscuration and have relatively older stellar populations when compared to the rest of the population. 
\item A fraction of about $\sim$5\% of the LBGs do have dust as evidenced by MIPS 24$mu$m detections, and are classified as ILLBGs. Those LBGs share many properties in common
 with the SMGs and preliminary results show that they can be detected at submm bands. It can therefore be suggested that a link between these two populations must exist.
\item Strong PAH features arising from the mid-infrared spectrum of an ILLBG at z=3.01 indicates the existence of dust in the interstellar medium of LBGs and suggest that
the emission is dominated by star formation rather than an AGN. 
\end{itemize}

\begin{figure}
\centering
 \includegraphics[width=10cm,height=4cm]{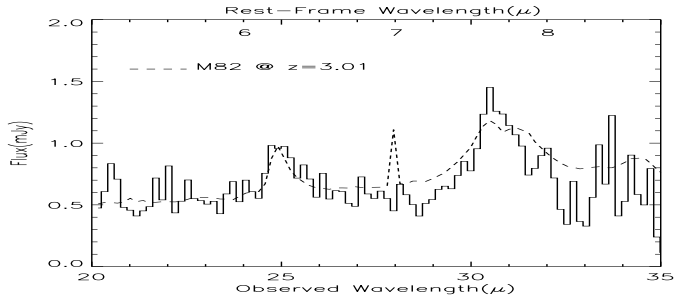}
  \caption {\small{IRS spectrum of EGS20 J1418+5236. The dashed line is the M82 SED shifted to $z=3.01$.
The spectrum has remarkably similar 6.2 and 7.7 $\mu$m PAH emission-feature strength and shape to those of M82.
We cross-correlated the IRS spectrum with that of M82 to derive a redshift of $z =3.01$$\pm$$0.016$.}}
\end{figure}

\end{document}